\documentclass[aps,amsmath,amssymb,prl,twocolumn,showpacs]{revtex4}
\usepackage{graphics}
\usepackage{array}

\def\d{{\rm d}}
\def\0{\varnothing}

\begin{document}
\title{Long-range attraction between probe particles mediated by a driven fluid}
\author{E. Levine$^{(a)}$, D. Mukamel$^{(a)}$, and G.M. Sch\"{u}tz$^{(b)}$}
\affiliation{$(a)$ Department of Physics of Complex Systems,
Weizmann Institute of Science, Rehovot, Israel 76100.\\
$(b)$Institut f\"{u}r Festk\"{o}rperforschung, Forschungszentrum
J\"{u}lich, 52425 J\"{u}lich, Germany.}

\date{\today}
\begin{abstract}
The effective interaction between two probe particles in a
one-dimensional driven system is studied. The analysis is carried
out using an asymmetric simple exclusion process with
nearest-neighbor interactions. It is found that the driven fluid
mediates an effective long-range attraction between the two
probes, with a force that decays at large distances $x$ as $-b/x$,
where $b$ is a function of the interaction parameters. Depending
on the amplitude $b$ the two probes may form one of three states:
(a) an unbound state, where the distance grows diffusively with
time; (b) a weakly bound state, in which the distance grows
sub-diffusively; and (c) a strongly bound state, where the average
distance stays finite in the long time limit. Similar results are
found for the behavior of any finite number of probes.
\end{abstract}

\maketitle

Probe particles are a powerful tool in the study of properties of
solutions. Interactions between solute particle arise from direct
interactions as well as a solvent-mediated part to which the free
energy of solvation contributes. The latter contribution is
sometimes rather significant, with surprising effects such as the
hydrophobic attraction \cite{BenN80} in which entropy plays the
dominant role. Strong entropic effects have also been observed in
anisotropic fluids \cite{aif}. In equilibrium one frequently
approximates these interactions by setting up a suitable model and
calculating an effective potential of mean force between solute
particles, using the potential distribution theorem \cite{Wido63}.
While more quantitative studies require quite sophisticated
modelling, simple lattice models often suffice to successfully
explain generic features of these interactions. Difficulties of
conceptual rather than practical nature, however, arise in
equilibrium-based approximation schemes if the solvent is in a
strongly non-equilibrium state and therefore no notion of free
energy exists. Analyzing the interactions between probe particles
in non-equilibrium fluids would thus be of great interest.

Far from equilibrium driven systems of particles moving steadily
have been a subject of extensive studies in recent years
\cite{Rev}. The minimal model which has been used to describe
these systems is the asymmetric simple exclusion process, whereby
self-avoiding particles hop on a lattice with rates which favor
motion along the direction of the drive \cite{ASEP,TASEP}. In one
dimension (1d) these studies resulted in detailed calculations of
a variety of steady state properties, including phase diagrams,
density correlation functions and other collective features. In
particular, it has been observed that a probe particle introduced
into a 1d driven fluid is attracted to regions of large density
gradient (on molecular scale) in the steady state density profile,
thus serving as a microscopic marker of shocks \cite{Shock}. A
local shock is accompanied by a long-range algebraically-decaying
density profile away from the probe. When the system contains two
probe particles, the density profile induces a long-range
attractive interaction between the probes. For example, in the
case of the totally asymmetric simple exclusion process (TASEP) on
a ring with two `second-class' probe particles, the steady-state
distribution of the distance $x$ between the two probes was found
to decay as $x^{-3/2}$ for large $x$ \cite{Derrida93}. This
implies that the two probes form a weakly bound state, where the
average distance is infinite. This exactly soluble case is very
specific though, as the second-class particles do not influence
the motion of the fluid particles. On the other hand, it is known
that the direct interaction between solute and solvent could have
a strong dynamical effects, and should not in general be ignored.
For example, direct interaction between solute and solvent may
lead to a non-monotonic dependence of solute and solvent diffusion
coefficients on the solute diameter \cite{Pere98}.

In this Letter we present a dynamical approach to the problem of
probe particles in a 1d driven fluid. This approach is applicable
to a broad class of 1d systems, which in addition to excluded
volume display short range interactions, and where
probes directly influence the motion of the fluid particles.
Moreover, it enables us to study not only the steady state
properties, but also the temporal relaxation to the steady state.
We find that within this broader class of systems, two probe
particles may form one of three states: unbound, weakly bound, or
a strongly bound state. In the unbound state the average distance
between the probe particles grows diffusively as $\sqrt{t}$, as is
the case for non-interacting particles. In the weakly bound state,
the steady-state distance distribution decays algebraically for
large $x$ with a power-law $x^{-\sigma}$, where the exponent
$1<\sigma<2$ is a function of the interaction parameters. The
steady state average distance is infinite. The approach to steady
state is sub-diffusive, whereby the average distance grows as
$t^{\nu}$ with $\nu<1/2$. In the strongly bound state the average
distance decays algebraically in time to a finite value at steady
state. In the steady-state, the distance distribution takes the
form $x^{-\sigma}$, with $\sigma>2$. The case of more than two
probe particles may also be studied within the approach suggested
in this Letter.

We now introduce a model within which the interaction between
probes may be analyzed. The model is defined on a 1d lattice of
ring geometry, where each site can be occupied by either a
positive particle $(+)$, a negative particle $(-)$, or it may be
occupied by a probe particle ($0$). Apart from the exclusion
interaction, the `charged' particles are subject to a
nearest-neighbor interaction, defined by the potential
\begin{equation} V = -
\frac{\epsilon}{4} \sum_i s_i s_{i+1}\;,
\end{equation}
where $s_i=0, \pm1$ according to the occupation of site $i$, and
$-0.8<\epsilon<1$ is the coupling constant \cite{AFM}. The
dynamics of the model
is defined by a random-sequential local dynamics, whereby a pair
of nearest-neighbor sites is selected at random, and the particles
are exchanged with rates
\begin{align}
\label{eq:model}
+- &\to -+ &\mbox{with rate}~1+\Delta V\nonumber\\
+0 &\to 0+\;,\;\;0- \to -0 &\mbox{with rate}~1\;.
\end{align}
Here $\Delta V$ is the difference in the potential $V$ between the
initial and final states. This dynamics conserves the number of
particles of each species, and is symmetric under the exchange of
charges and spatial direction. The model is a generalization of
\cite{KLS}, which has been studied in \cite{KLSZRP,Evans04}. In
the case of no nearest neighbor interaction, $\epsilon=0$, the
dynamics defined above reduces to that of the TASEP with
second-class particles \cite{Shock,Derrida93} which are the
probes.

In the case of no probe particles, the steady state of this model
has an Ising-like measure \cite{KLS}. This is also expected to be
the local steady-state measure away from any probe when the
density of probes is zero. We consider first the case of two probe
particles in the system. Our aim is to derive dynamical equations
for the two probes, and study the evolution in time of the
distance $n$ between them, starting from $n=0$.

\begin{figure}
\centerline{\resizebox{0.4\textwidth}{!}{\includegraphics{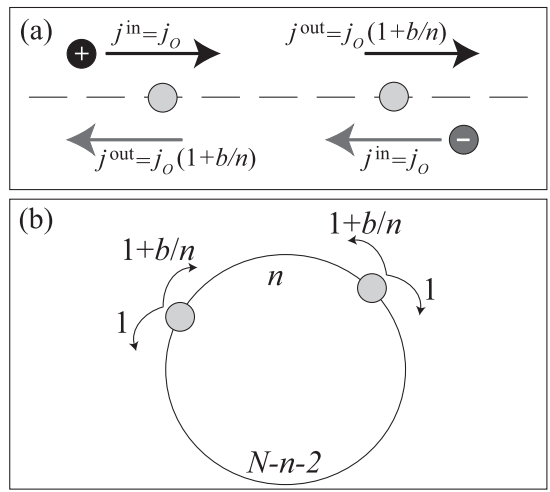}}}
\caption{(a) Schematic diagram of the positive (black) and
negative (gray) currents, as seen from the two probe particles.
Here $n$ is the distance between the two probes. For $\epsilon=0$
one has $j_\circ=1/4$ and $b=3/2$. (b) A random-walk
representation of the dynamics of the two probe particles, as
deduced from the currents in (a). Each probe carries out forward
and backward hops with the indicated rates.} \label{fig:toy}
\end{figure}

For $\epsilon=0$ it has been shown \cite{Derrida93} that the
statistical weight of all configurations with a given $n$ is
proportional to $Z_n$, the partition function of the TASEP on an
open chain of length $n$ in the maximal current phase
\cite{TASEP}. This observation has been used to show that the
probability to find the two probes at a distance $n$ from each
other decays for large $n$ as $n^{-3/2}$ \cite{Derrida93}. In
addition, this result can be used to estimate the currents of
particles which go in and out of the segment trapped between the
two probe particles. One finds that the outgoing current of $+$
($-$) particles through the right (left) probe takes the form of
the steady state current of the TASEP, given for large $n$ by
$j^{\rm out}_n=\frac{1}{4}\left(1+\frac{3/2}n\right)$. The
opposing currents, namely that of $-$ ($+$) particles incoming
through the right (left) probe, take the form $j^{\rm
in}=\frac{1}{4}\left(1+\frac{3/2}{N-n-2}\right)$. In the limit we
are concerned with, namely $N \gg n$ and $N \to \infty$, this
current is well approximated by $j^{\rm in}=\frac{1}{4}$
[Fig.~\ref{fig:toy}(a)].

These considerations have been extended to
the case $\epsilon \neq 0$ \cite{KLSZRP}. Following
\cite{KLSZRP} one expects
that the current of $+$ ($-$) particles bypassing the
right (left) probe takes the same form as that of the current in
an open segment of the same length, governed by the same dynamics.
This current is given by $j^{\rm
out}_n=j_\circ(\epsilon)\left(1+b(\epsilon)/n\right)$, where
\begin{gather}
\label{eq:j} j_\circ(\epsilon) = \frac{\upsilon +
\epsilon}{\upsilon^{3}} \;,\quad
b(\epsilon)=\frac{3}{2}\;
\frac{(2+\epsilon)\upsilon+2\epsilon}{2(\upsilon+\epsilon)}\;,
\end{gather}
and $ \upsilon=\sqrt{\frac{1+\epsilon}{1-\epsilon}}+1$. For the
relevant values of $\epsilon$ one has $0\leq b\leq9/4$. Also,
similar to the $\epsilon=0$ case, the incoming current is
given in the large $N$ limit by $j^{\rm in}=j_\circ(\epsilon)$.

The considerations described above may be used to derive dynamical
equations for the two probes. These considerations suggest that
the two probe particles behave as two coupled random walkers. Each
probe moves away from the other probe with rate $j_\circ$, and
move towards it with rate $j_\circ\left(1+b/n\right)$
[Fig.~\ref{fig:toy}(b)]. Thus, one can model the time evolution of
the distance $n$ between the probes by the master equation
\begin{subequations}\label{eq:master}\begin{multline}
\frac{\partial P(n,t)}{\partial t} =
P(n-1,t)-P(n,t)\\+\left(1+\frac{b}{n+1}\right)P(n+1,t)-
\left(1+\frac{b}{n}\right)P(n,t)
\end{multline}
for $n>0$, and
\begin{equation}
\frac{\partial P(0,t)}{\partial t} =
-P(0,t)\\+\left(1+b\right)P(1,t)
\end{equation}
\end{subequations} for $n=0$. Here the time $t$ is a rescaled by a
factor $2j_\circ$. For large $n$ one can use the continuum limit,
and~\eqref{eq:master} can be rewritten as a Fokker-Planck equation
(FPE)
\begin{equation}
\label{eq:FPE} \frac{\partial P(x,t)}{\partial t} =
\frac{\partial^2 P(x,t)}{\partial x^2} +\frac{\partial}{\partial
x}\left(\frac{b}{x}P(x,t)\right)\;,
\end{equation}
with the boundary conditions $P(x,t)\to0$ as $x\to\infty$. It is
easy to see that the steady-state solution of this equation is of
the form
\begin{equation}
P_{\mbox{st}}(x) \sim x^{-b}\;.
\end{equation}
This in agreement with the exact result $x^{-3/2}$
in the case $\epsilon=0$ \cite{Derrida93}. 
Note that the force term  $b/x$ in the FPE (\ref{eq:FPE})
corresponds to a logarithmically increasing fluid-mediated
effective potential $U \sim \ln{x}$ between the probe particles.

In order to study the approach to steady state we make the scaling
hypothesis $P(x,t) = x^{-b}t^{-\beta}f(x/t^{\alpha})$, with $f(u)$
a scaling function, which satisfies the boundary condition $f(0)=
\mbox{const.}$,  and the normalization condition
$\int_\theta^\infty P(x)\,\d x = 1$. Here a cutoff $\theta$ is
introduced to prevent divergence at $x\to 0$ for $b\geq1$.
Substituting this form into~\eqref{eq:FPE} one finds that a
non-trivial solution exists only when $\alpha=1/2$. In the range
$b>1$ the normalization condition yields $\beta=0$. Defining
$u=x/\sqrt{t}$, the scaling function $f(u)$ satisfies for $b>1$
\begin{equation}
\left(\frac{1}{2}u^2-b\right)f'(u)+uf''(u)=0.
\end{equation}
This equation reduces to $f'(u)\sim u^b\exp(-u^2/4)$, and its
solution is given by $f(u) \sim
\gamma\left(\frac{b+1}{2},\frac{u^2}{4}\right)$, where
$\gamma(a,x)$ is the incomplete $\Gamma$-function. In the range
$b<1$ the normalization condition yields $\beta=(1-b)/2$ and hence
\begin{equation}
\left(\frac{1}{2}u^2-b\right)f'(u)+uf''(u)=-\frac{1-b}{2}u f(u)
\end{equation}
which is solved by $f(u) \sim \exp(-u^2/4)$.

We now turn to calculate the average distance between the
particles, $ \left<x(t)\right> = {\int_\theta^\infty\d x\,x
P(x,t)}$. Inspecting the large $t$ behavior of this integral one
finds
\begin{equation}
\label{eq:nu} \left<x(t)\right> \sim
\begin{cases}
t^{1/2}&b<1\\
t^{1/2}/\log(t)&b=1\\
t^{1-b/2}&1<b<2\;.\\
\log(t)&b=2\\
A+B t^{-(b/2-1)}&b>2
\end{cases}
\end{equation}
Here $A$ and $B$ are some non-universal constants. It is evident
that depending on the value of $b$ three regimes can be
identified. For $b<1$ the two probes behave as two decoupled
particles, and the distance between them grows diffusively. For
$1\leq b\leq2$ the probes are weakly bound, and the distance
between them diverges sub-diffusively. For $b>2$ the two probes
are strongly bound, and the average distance between them
approaches a constant algebraically in time.

\begin{figure}
\centerline{\resizebox{0.5\textwidth}{!}{\includegraphics{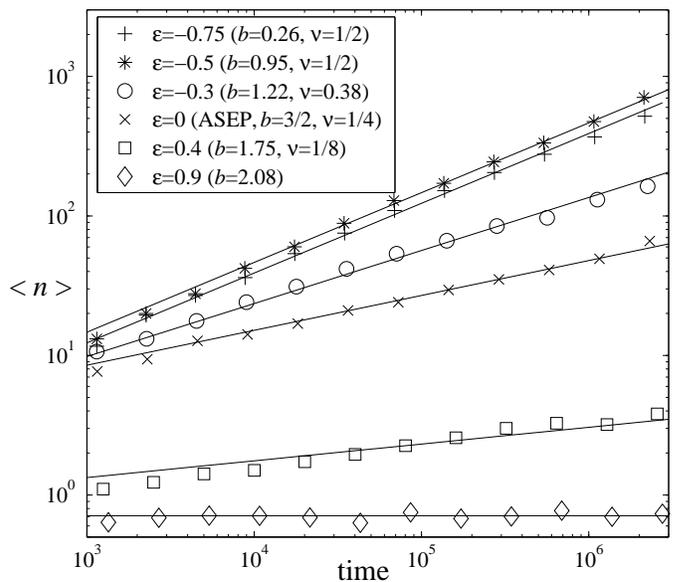}}}
\caption{Results of numerical simulations for the average distance
between the two probes for several values of $\epsilon$, and thus
of $b$. Lines are drawn with the expected slope, according to
Eq.~\ref{eq:nu}.} \label{fig:nu}
\end{figure}

To test the validity of these results we carried out numerical
simulations of the model with two probe particles. The simulations
were performed on a ring of length $N=10^5$, with equal number of
$+$ and $-$ particles. Starting with the two probes located on
nearest neighbor sites, we followed the evolution of their
distance $n(t)$. The average distance $\left<n(t)\right>$ was
obtained by averaging over $5000$ realizations of the noise.
Results for different values of $b$ in the three regimes are
presented in Fig.~\ref{fig:nu}. These results agree very well with
the theoretical predictions.

We now consider the case of a finite number $M$ of probes, and
show that the scaling behavior, Eq.~\ref{eq:nu}, holds for the
average distance between any pair of probes. The evolution of $M$
probes can be be modelled as a zero-range process (ZRP)
\cite{ZRP}, defined on a one-dimensional ring of $M$ sites, where
each site can be occupied by any number of particles.
At any time step a site is
chosen at random, and one of its particles hops to either one of
its two nearest neighbors with the same rate $w_n =
j_\circ\left(1+b/n\right)$. Here $n$ is the occupation of the
departure site.  To make the correspondence between this ZRP and
the driven model \eqref{eq:model} we identify the occupation
number of site $i$ with the distance between the probes $i$ and
$i+1$ in the driven model. Moreover, since we are interested in
the limit $N\to\infty$, we take site $i=M$ to be occupied by an
infinite number of particles. This site serves as a reservoir,
omitting particles with current $j_\circ$ to both sides. For this
zero-range process it can be shown \cite{us} that the distribution
of occupation numbers in the steady state takes the form of a
product measure,
\begin{equation}
P_{\mbox{st}}\left(n_1,n_2,\cdots,n_{M-1}\right) \sim
\prod_{i=1}^{M-1}n_i^{-b}\;.
\end{equation}
In the particle system, this steady-state distribution holds
exactly only in the case $\epsilon=0$. However, it has been shown
that the correlations between adjacent segments can be ignored at
large distances $n_i$ also for $\epsilon \neq0$
\cite{KLSZRP,Evans04}. With the steady-state distribution at hand,
the temporal approach to steady-state can be studied in a similar
manner to the case of two probes. In particular, we make the
scaling hypothesis
\begin{multline}
P\left(x_1,\cdots,x_{M-1},t\right) = \\x_1^{-b}\cdots
x_{M-1}^{-b}t^{-(M-1)\beta}f\left(\frac{x_1}{\sqrt{t}},\cdots,\frac{x_{M-1}}{\sqrt{t}}\right)\;,
\end{multline}
where, again, $x_i$ is the continuous variable corresponding to
$n_i$. A straightforward analysis of this scaling form shows that
\eqref{eq:nu} holds for all $\left<x_i\right>$, and thus for the
distance between any pair of probes. This agrees well with
numerical simulations of systems with $5$ probes, supporting the
scaling hypothesis (details are not presented).

Our discussion so far was limited to the case where the number of
positive particles was equal the number of negative particles. In
\cite{Evans04} it has been argued that the physical picture of
domain currents underlying the dynamical approach developed above
still holds when the densities of the two species are not equal.
In this case, the amplitude $b$ becomes a function of both
$\epsilon$ and $\eta$, where $\eta$ is the density of, say, the
positive particles (see \cite{Evans04} for explicit expressions).
It is evident that $b(\epsilon,\eta)$ increases when $\eta$
deviates from $\frac12$. Thus, increasing the density of one
species at the expense of the other results in larger $b$, and may
change the state of the probes from unbound to weakly bound to
strongly bound state.

In summary, we introduced in this Letter a dynamical approach for
the study of probe particles in one-dimensional driven fluids.
This approach generalizes the exact result of \cite{Derrida93},
obtained for the steady-state distribution of two probe particles
which do not affect the motion of the fluid, to a broader class of
models, which are not exactly solvable. Here any finite number of
probe particles, which influence the motion of the fluid, are
considered. Moreover, the dynamical approach enables one to study
not only steady-state properties, but also the long-time temporal
behavior of the probes. It is found that a logarithmically
increasing fluid-mediated effective potential $U \sim \ln{x}$ acts
between the probes when they are a distance $x$ apart. It is
remarkable that the probe particles behave as if they were in
thermal equilibrium even though the fluid is very far from
equilibrium. This is reminiscent of the motion of a
boundary-induced shock in non-conserving driven systems in contact
with a reservoir. The position of the shock has recently been
found to perform Brownian motion in an effective potential
dynamically generated by the fluid medium \cite{Rako03}.

\begin{acknowledgments} G.M.S. thanks the Weizmann Institute for
kind hospitality. The support of the Albert Einstein Minerva
Center for Theoretical Physics, Israel Science Foundation, and
Deutsche Forschungsgemeinschaft (grant Schu827/4), is gratefully
acknowledged.
\end{acknowledgments}

\end{document}